\pdfoutput=1

\documentclass[11pt]{article}

\usepackage[final]{acl}

\usepackage{times}
\usepackage{latexsym}

\usepackage[T1]{fontenc}

\usepackage[utf8]{inputenc}

\usepackage{microtype}

\usepackage{inconsolata}

\usepackage{graphicx}

\usepackage{multirow}
\usepackage{tabularx}
\usepackage{booktabs}
\usepackage{CJKutf8}

\title{ChildMandarin: A Comprehensive Mandarin Speech Dataset \\
for Young Children Aged 3-5}

\author{
 \textbf{Jiaming Zhou\textsuperscript{1}}, \textbf{Shiyao Wang\textsuperscript{1}},  \textbf{Shiwan Zhao\textsuperscript{1}}, \textbf{Jiabei He\textsuperscript{1}}, \textbf{Haoqin Sun\textsuperscript{1}}, \textbf{Hui Wang\textsuperscript{1}}, 
 \\
   \textbf{Cheng Liu\textsuperscript{1}}, \textbf{Aobo Kong\textsuperscript{1}}, \textbf{Yujie Guo\textsuperscript{1}}, \textbf{Xi Yang\textsuperscript{2}},  \textbf{Yequan Wang\textsuperscript{2}}, \textbf{Yonghua Lin\textsuperscript{2}}, \textbf{Yong Qin\textsuperscript{1}}\thanks{Yong Qin is the corresponding author.} 
\\
\\
     \textsuperscript{1} College of Computer Science, Nankai University, \\
     \textsuperscript{2} Beijing Academy of Artificial Intelligence, Beijing, China,
\\
 \small{
   \textbf{Correspondence:} \href{mailto:zhoujiaming@mail.nankai.edu.cn}{zhoujiaming@mail.nankai.edu.cn}, \href{mailto:qinyong@nankai.edu.cn}{qinyong@nankai.edu.cn}
 }
}

\begin{document}
\maketitle
\begin{abstract}

Automatic speech recognition (ASR) systems have advanced significantly with models like Whisper, Conformer, and self-supervised frameworks such as Wav2vec 2.0 and HuBERT. However, developing robust ASR models for young children's speech remains challenging due to differences in pronunciation, tone, and pace compared to adult speech. In this paper, we introduce a new Mandarin speech dataset focused on children aged 3 to 5, addressing the scarcity of resources in this area. The dataset comprises 41.25 hours of speech with carefully crafted manual transcriptions, collected from 397 speakers across various provinces in China, with balanced gender representation. We provide a comprehensive analysis of speaker demographics, speech duration distribution and geographic coverage. Additionally, we evaluate ASR performance on models trained from scratch, such as Conformer, as well as fine-tuned pre-trained models like HuBERT and Whisper, where fine-tuning demonstrates significant performance improvements. Furthermore, we assess speaker verification (SV) on our dataset, showing that, despite the challenges posed by the unique vocal characteristics of young children, the dataset effectively supports both ASR and SV tasks. This dataset is a valuable contribution to Mandarin child speech research. The dataset is now open-source and freely available for all academic purposes on \url{https://github.com/flageval-baai/ChildMandarin}.
\end{abstract}

\begin{table*}[t]
\small
\centering
\begin{tabular}{cccccccc}
\toprule
Corpus               & Age range & \# Speakers & Dur. (hrs) & Style & Year & Trans. & Avail. \\
\midrule
Tong Corpus         & 1;7-3;4     & 1           & 22       & Interactions        & 2018 & Y             & Y            \\
CASS CHILD           & 1-4         & 23          & 631      & Spontaneous speech & 2012 & P             & N            \\
SLT-CSRC C1          & 7-11        & 927         & 28.6     & Reading             & 2021 & Y             & N            \\
SLT-CSRC C2         & 4-11        & 54          & 29.5     & Conversation        & 2021 & Y             & N            \\
SingaKids   & 7-12        & 255         & 75      & Reading             & 2016 & Y             & Y  \\
Ours                & 3-5          & 397         & 41.3        & Conversation & 2024 &Y              &Y               \\
\bottomrule
\end{tabular}
\caption{Summary of Chinese child speech datasets: age range, speaker count, duration, and availability. Dur.: duration. Trans.: transcriptions (P: partial). Avail.: availability.}
\label{chinese_dataset}
\end{table*}

\begin{table*}[t]
\centering
\small
\begin{tabularx}{\textwidth}{Xccccc}
\toprule
Corpus              & Language & Age range & \# Speakers & Dur.(hrs) & Year \\
\midrule
Providence Corpus~\citep{demuth2006word} & English  & 1-3     & 6    & 363      & 2006 \\
Lyon Corpus~\citep{demuth2008prosodically} & English  & 1-3     & 4    & 185      & 2008 \\
TBALL~\citep{kazemzadeh2005tball} & English  & K - G4  & 256  & 40       & 2005 \\
CU Children's Read and Prompted Speech Corpus~\citep{hagen2003children} & English  & K - G5  & 663  & -        & 2003 \\
CSLU Kids' Speech Corpus~\citep{cslu} & English  & K-G10   & 1,100 & -        & 2007 \\
CU Story Corpus~\citep{hagen2003children} & English  & G3-G5 & 106  & 40       & 2003 \\
MyST Corpus~\citep{pradhan2024my}& English  & G3-G5   & 1,371 & 393      & 2024 \\
PF-STAR Children's Speech Corpus~\citep{inproceedings}& English  & 4-14    & 158  & 14.5     & 2005 \\
The CMU Kids Corpus~\citep{eskenazi1997cmu}& English  & 6-11    & 76   & -        & 1997 \\
TIDIGITS~\citep{Gary1993s10}& English  & 6-15    & 101  & -        & 1993 \\
CID children’s speech corpus~\citep{lee1999acoustics}& English  & 5-18    & 436  & -        & 1999 \\
Speechocean762~\citep{zhang2021speechocean762}& English  & 5-18    & 125  & 6        & 2021 \\
Non-Native children’s speech corpus~\citep{radha2022audio}& English  & 7-12    & 20   & 3.3     & 2022 \\
Demuth Sesotho Corpus~\citep{demuth1992acquisition}& Sesotho  & 2-4     & 59   & 98       & 1992 \\
CHIEDE~\citep{garrote2008chiede}& Spanish  & 3-6     & 59   & $\sim$8  & 2008 \\
IESC-Child~\citep{perez2020iesc}& Spanish  & 6-11    & 174  & $\sim$35 & 2020 \\
JASMIN-CGN Corpus~\citep{cucchiarini2008recording}& Dutch    & 7-16    & -    & $\sim$64 & 2008 \\
SANACS~\citep{kruyt2024slovak}& Slovak   & 6-12    & 67   & $\sim$15 & 2024 \\
CFSC~\citep{pascual2012developing}& Filipino & 6-11    & 57   & $\sim$8  & 2012 \\
Swedish NICE Corpus~\citep{bell2005swedish}& Swedish  & 8-15    & 5,580 & $\sim$6        & 2005 \\
\bottomrule
\end{tabularx}
\caption{Summary of child speech datasets in other languages, where K denotes kindergarten while G denotes grade.}
\label{dataset_other_lang}
\end{table*}

\section{Introduction}

Automatic Speech Recognition (ASR) technology has become increasingly prevalent across various applications, ranging from virtual assistants and educational tools to accessibility services for individuals with disabilities \citep{kennedy2017child}. In particular, child speech recognition holds great potential in educational settings, such as language learning applications, reading tutors, and interactive systems. However, despite the rapid advancements in ASR technology, the performance of most systems—whether state-of-the-art or commercial—remains suboptimal when applied to children's speech \citep{fan24b_interspeech}.

ASR systems are predominantly trained on adult speech \cite{zhou2024knn}, making them highly effective for everyday interactions but ill-suited for children due to physiological differences in vocal tract development, higher pitch, and inconsistent pronunciation \citep{lee1997analysis,gerosa2009review}. Children's speech also exhibits considerable variability in articulation, speech patterns, and vocabulary, further complicating the recognition process \citep{benzeghiba2007automatic,bhardwaj2022automatic}. These challenges are compounded by the lack of sufficient child-specific training data, which is crucial for developing ASR systems that can accurately and reliably understand children's speech across different age groups. However, datasets focused on young children are extremely rare \citep{graave24_interspeech}. Most existing speech datasets either concentrate on adult speakers or cover older children, overlooking the unique linguistic and developmental characteristics of younger children. This gap is critical, as the scarcity of training data limits the ability of ASR systems to perform well on speech from this age group \cite{zhou2023madi}.

 Although there are a few open-source Mandarin speech datasets for children \citep{tong_corpus,cass_child,slt_csrc,singakids}, they are often limited in scope. For instance, the Tong Corpus \citep{tong_corpus} records the speech of a single child from ages 1;7 to 3;4, which is useful for certain research areas, but insufficient for ASR development due to the lack of speaker diversity. Similarly, while the CASS CHILD corpus \citep{cass_child} includes data from 23 children aged 1 to 4 years, a portion of 80 hours is transcribed, it is not publicly available, restricting its use in ASR research.
Children’s speech poses unique challenges, with frequent mispronunciations, ungrammatical expressions, and child-specific vocabulary. To address these issues, it is essential to collect data from a large number of speakers, ensuring substantial amounts of data per speaker to capture linguistic variability and improve the generalization of ASR models. Existing datasets, such as the SingaKids-Mandarin \citep{singakids} and SLT-CSRC \citep{slt_csrc}, primarily focus on older children (aged 7-12), leaving a gap for younger age groups.

Constructing a dedicated speech dataset for young children is crucial. It addresses a significant gap in existing resources and provides a foundation for developing ASR systems specifically tailored to young children. In this paper, we introduce a Mandarin speech dataset designed for children aged 3 to 5, comprising 41.25 hours of speech from 397 speakers across 22 of China’s 34 provincial-level administrative divisions. Our evaluations of ASR models and speaker verification (SV) tasks demonstrate substantial improvements, underscoring the dataset's effectiveness in advancing technology for children's speech. This dataset bridges the gap in age-specific speech data by incorporating a wide range of speakers and extensive regional diversity. It represents a valuable contribution to Mandarin child speech research and holds significant potential for applications in educational technology and child-computer interaction.

\section{Related Work} \subsection{Child Speech Recognition Corpora in Mandarin Chinese}

Publicly available child speech corpora for Mandarin Chinese are highly limited, particularly for younger age groups, as shown in Table \ref{chinese_dataset}. The few existing datasets are either too small in terms of speakers or lack accessibility, which restricts their utility for developing robust ASR systems.

The Tong Corpus \citep{tong_corpus} is a longitudinal dataset that records the speech of a single child, Tong, with one hour of recordings per week from ages 1;7 to 3;4. Although this corpus is valuable for research on language acquisition, its use in ASR development is limited by its single-speaker nature, which cannot provide the diversity needed for model generalization.

\citet{cass_child} collected the CASS CHILD dataset, which contains 631 hours of speech from 23 children aged 1 to 4 years. However, only about 80 hours of this dataset are labeled with transcriptions, and, critically, the dataset is not publicly accessible. This restricts its use in ASR experiments and highlights the difficulty of obtaining child speech corpora in Mandarin.

The SingaKids-Mandarin Corpus \citep{singakids} contains 75 hours of speech data from 255 children aged 7 to 12, which is suitable for ASR training. This corpus encompasses diverse linguistic contexts. However, it focuses exclusively on children aged 7 to 12 and does not address the speech of younger children, which represents a significant gap in Mandarin ASR research.

Another important dataset is SLT-CSRC \citep{slt_csrc}, which consists of two collections: SLT-CSRC C1 and C2. The former includes 28.6 hours of reading-style speech from 927 children aged 7 to 11, while the latter consists of 29.5 hours of conversational speech from 54 children aged 4 to 11. Although these datasets provide valuable speech data for Mandarin ASR, they were only available for participants of the SLT 2021 challenge and are no longer publicly accessible.

In summary, for Mandarin child speech, only the Tong Corpus and SingaKids-Mandarin datasets are available upon request, and both are limited in terms of speaker diversity and age range coverage. This lack of publicly accessible child speech corpora, particularly for younger children, continues to be a significant challenge in Mandarin ASR development.

\subsection{Child Speech Corpora in Other Languages}

In other languages, especially English, a wider variety of child speech corpora exists, as shown in Table \ref{dataset_other_lang}. These corpora differ significantly in size, age range, and speaker diversity, reflecting various research priorities. 
However, many still lack sufficient coverage for younger children, a crucial age group for advancing ASR development.

English corpora, in particular, are among the most well-represented. For example, the Providence~\citep{demuth2006word} and Lyon Corpora~\citep{demuth2008prosodically} focus on early childhood speech (ages 1-3), offering 363 and 185 hours of recordings, respectively. Despite their extensive durations, these datasets are limited in the number of speakers, with only 6 and 4 children represented, respectively. On the other hand, larger datasets such as the MyST Corpus~\citep{pradhan2024my} offer 393 hours of conversational speech from virtual tutoring sessions in elementary school science, collected from 1,371 children in grades 3 to 5. This broader speaker diversity is highly advantageous for training robust ASR systems.

Other notable English datasets include the CSLU Kids' Speech Corpus~\citep{cslu}, which features reading recordings from over 1,100 children from kindergarten through grade 10 including simple words,digits and sentences, and the TBALL Corpus~\citep{kazemzadeh2005tball}, which contains speech from 256 children in kindergarten through grade 4. These datasets contribute valuable resources for developing ASR systems for various childhood age ranges and linguistic styles.

Child speech datasets in other languages are less common and typically smaller. For example, the Demuth Sesotho Corpus~\citep{demuth1992acquisition} offers 98 hours of speech from 59 children aged 2 to 4, focusing on a non-Indo-European language, while the CHIEDE corpus~\citep{garrote2008chiede} contains around 8 hours of speech from 59 Spanish-speaking children aged 3 to 6. The IESC-Child Corpus~\citep{perez2020iesc} provides about 35 hours of Spanish speech from 174 children aged 6 to 11.

For European languages, the JASMIN-CGN Corpus~\citep{cucchiarini2008recording} offers 64 hours of Dutch speech from children aged 7 to 16, and the Swedish NICE Corpus~\citep{bell2005swedish} features data from 5,580 children aged 8 to 15. Although the NICE Corpus stands out for its large number of speakers, the total duration of recordings is relatively short, and similar limitations regarding younger children persist across these corpora.

Although these corpora are valuable, they reveal a significant shortage of publicly accessible child speech datasets for many languages, particularly for younger children and non-European languages. This gap underscores the urgent need for diverse, well-annotated child speech corpora to support ASR systems capable of generalizing across different languages, age ranges, and regions.

Our Mandarin Chinese dataset alleviates this gap by focusing on children aged 3 to 5, a critical yet underrepresented age group in ASR research. With 397 speakers and 41.25 hours of diverse, geographically distributed speech data, it offers a significant contribution to the field, especially given the scarcity of similar datasets for young children in non-European languages.

\begin{table}[t]
\centering
\begin{tabular}{ccccc}
\toprule
Split & \# Spk. & \# Utt. & Dur. (hrs) & Avg. (s) \\ \midrule
Train & 317     & 32,658     & 33.35            & 3.68     \\
Dev   & 39      & 4,057      & 3.78             & 3.35     \\
Test  & 41      & 4,198      & 4.12             & 3.53     \\
Sum   & 397     & 40,913     & 41.25            & 3.52    \\ \bottomrule
\end{tabular}
\caption{Summary of dataset splits, including the number of speakers (\# Spk.) and utterances (\# Utt.), total duration (Dur.), and average utterance length (Avg.).}
\label{dataset_detail}
\end{table}

\begin{figure}[!t]
  \centering
  \includegraphics[width=0.9\linewidth]
  {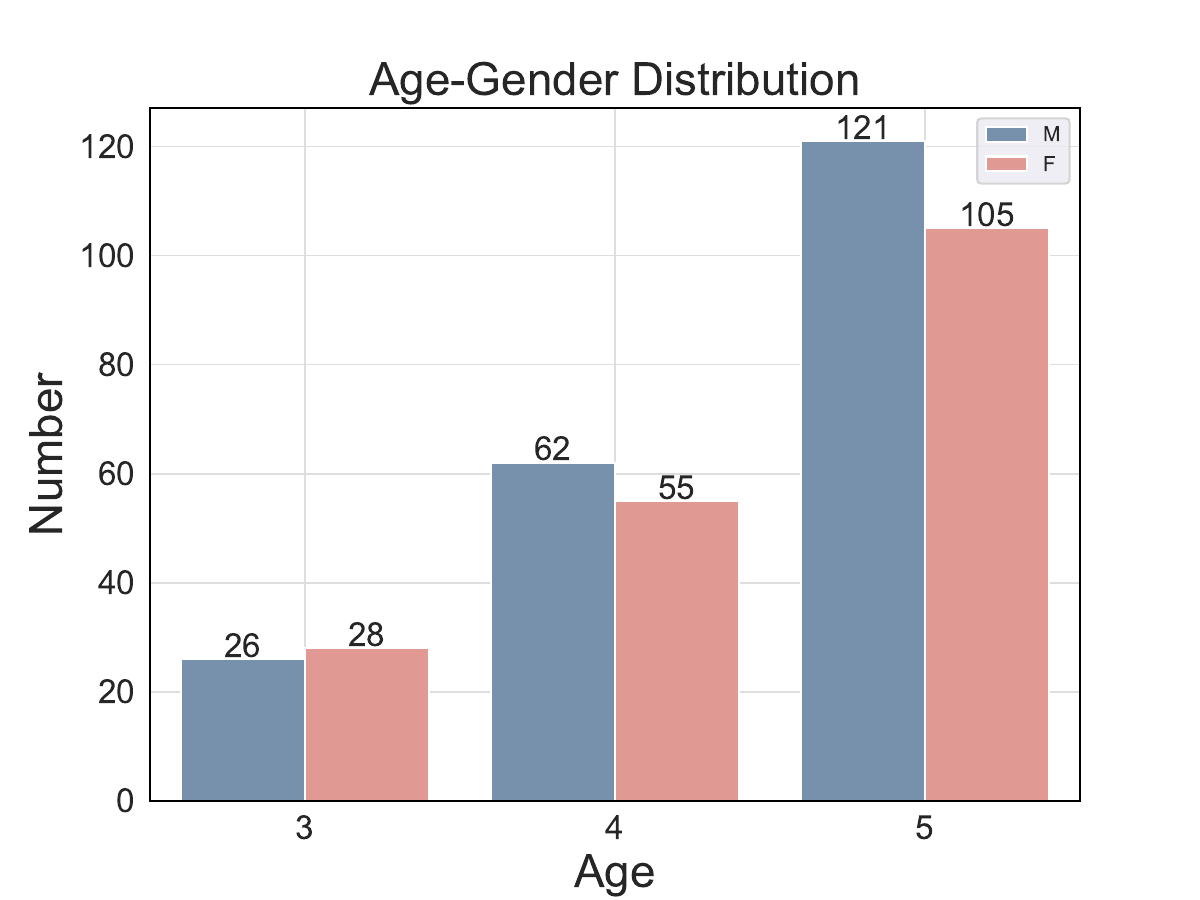}
  \caption{Distribution of speakers by age and gender in our dataset}
  \label{age_gender}
\end{figure}

\section{Dataset description}
\subsection{Dataset details}

The dataset consists of 41.25 hours of speech data with carefully crafted character-level manual transcriptions, collected from Mandarin-speaking children aged 3 to 5 years. The gender distribution is balanced across all age groups. To ensure geographic coverage, speakers were selected from different regions of China. A total of 397 speakers participated, representing 22 out of 34 provincial-level administrative divisions. Accents were classified into three categories: heavy (H), moderate (M), and light (L).

Our data collection occurred in a conversational context to promote natural interaction, with parents present throughout the sessions to provide emotional comfort and support for the children.  
The recording content was unrestricted, focusing on age-appropriate daily communication, ensuring that children engaged in familiar and non-stressful activities.

Prior to data collection, informed consent was obtained from the parents or legal guardians of all participants.
The consent process included detailed explanations of the study's purpose, procedures, and the intended use of the data for academic research.  Parents were explicitly informed about their right to withdraw consent at any time without any repercussions.

All recordings followed standardized collection and annotation protocols. Speech samples were captured using smartphones, with a nearly even split between Android (216) and iPhone (181) devices. Each session took place in quiet indoor environments, with minimal background noise tolerated due to the young age of participants. The recordings were in WAV PCM format, with a 16kHz sampling rate and 16-bit precision, ensuring high-quality audio without clipping or volume inconsistencies. Silence segments of approximately 0.3 seconds were preserved at the beginning and end of each valid speech segment, and utterances containing fewer than three characters were excluded.

Character-level manual annotations were performed by professional transcribers, who meticulously adhered to the audio content, including stutters, disfluencies, and developmental speech patterns. Regional pronunciation variations were transcribed faithfully. Additionally, numbers were transcribed as pronounced, maintaining consistency with the intended meaning of the speech.

\subsection{Statistics}

\begin{figure}[!t]
  \centering
  \includegraphics[width=1.0\linewidth]
  {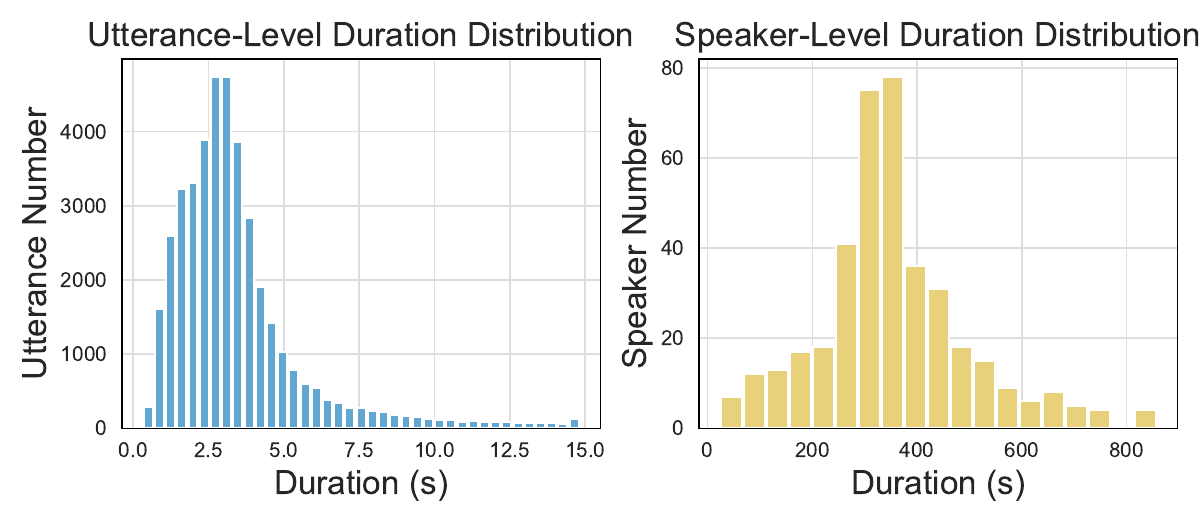}
  \caption{Utterance-level and speaker-level duration distribution in our dataset}
  \label{duration}
\end{figure}

\begin{figure}[!t]
  \centering
  \includegraphics[width=1.0\linewidth]
  {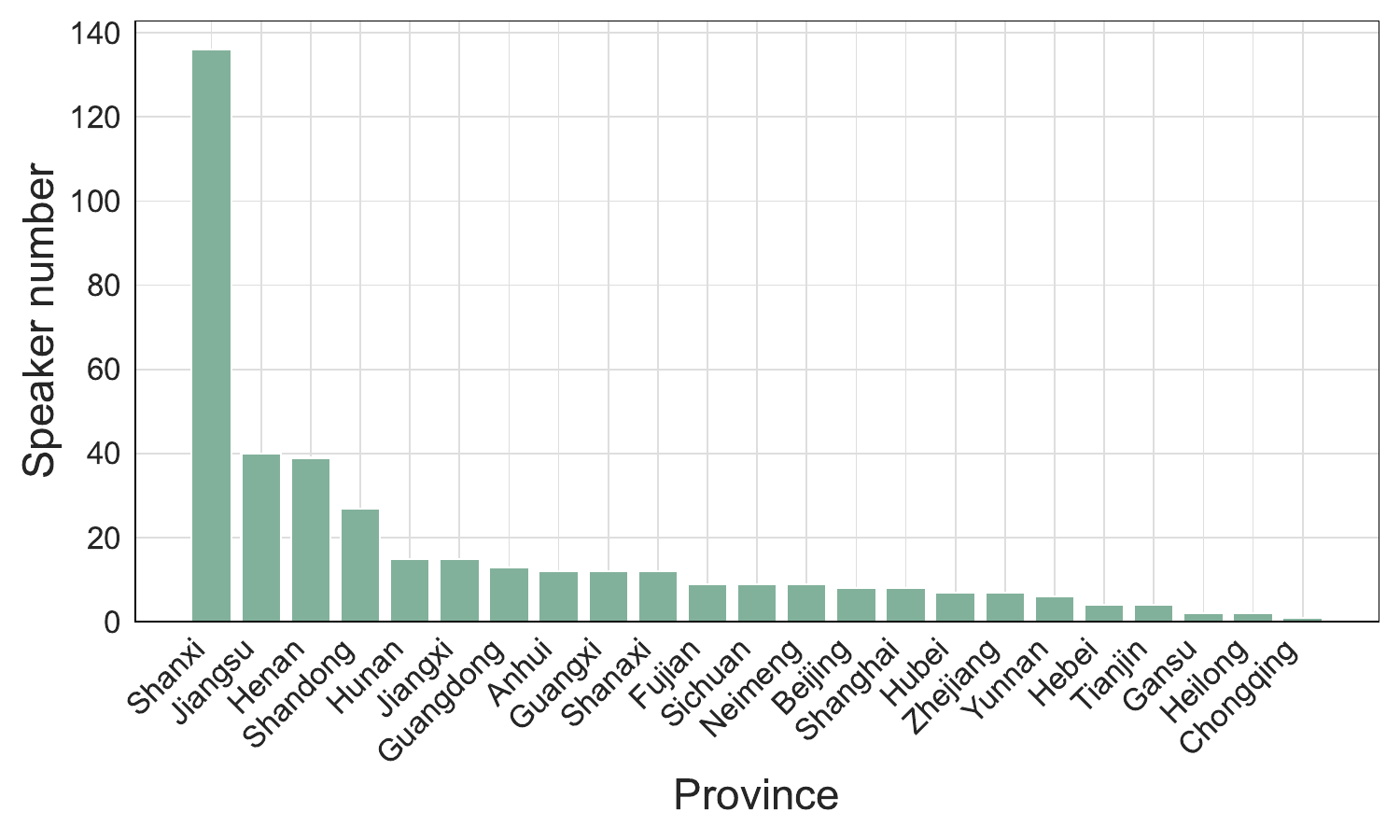}
  \caption{Geographic distribution of speakers in our dataset}
  \label{province_distribution}
\end{figure}

As shown in Table \ref{dataset_detail}, our dataset consists of three subsets: training (317 speakers), validation (39 speakers), and test (41 speakers), with no overlap between speakers across the subsets. We further analyze the distribution of speakers based on age, gender, birthplace, accent and recording device.

The age and gender distribution in the dataset, depicted in Figure \ref{age_gender}, highlights a decrease in the number of speakers as age decreases, which reflects the challenges in recruiting younger participants. Despite this, the gender distribution remains balanced across all age groups.

The distribution of utterance lengths and total speaking duration per speaker is presented in Figure \ref{duration}. Most utterances are between 1 and 5 seconds long, with very few exceeding 10 seconds. Additionally, the majority of speakers have a total speaking duration between 200 and 600 seconds, which is essential for developing ASR systems tailored to young children.

The geographic distribution of speakers, spanning 22 of China's 34 provincial-level administrative divisions, is summarized in Figure \ref{province_distribution}. Despite recruitment challenges, broad regional representation was achieved, with Shanxi contributing the highest number of participants (136), followed by Jiangsu (40) and Henan (39). Provinces such as Shaanxi, Shandong, and Hunan also contribute significantly. Although some regions, including Gansu, Heilongjiang, and Chongqing, have fewer participants, their inclusion enhances the dataset’s comprehensive geographic coverage.

\begin{figure}[!t]
  \centering
  \includegraphics[width=0.9\linewidth]{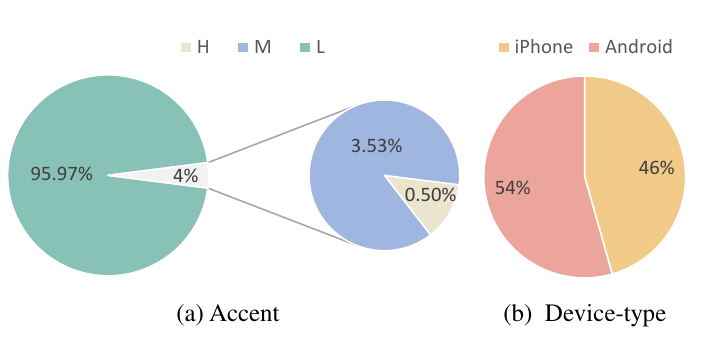}
  \caption{Proportions of accents and recording devices in our dataset}
   \label{accent_device}
\end{figure}

Speaker accents and recording devices are analyzed in Figure \ref{accent_device}. Accents are categorized into three levels: heavy (H), moderate (M), and light (L), with the majority of speakers exhibiting light accent variation. Only around 4\% of speakers are categorized as having moderate or heavy accents. Furthermore, a balanced representation of iPhone and Android devices was achieved to support diverse ASR system requirements.

\begin{table*}[t]
\small
\centering
\begin{tabular}{ccccccc}
\toprule
\multirow{2}{*}{Encoder} & \multirow{2}{*}{Loss} & \multirow{2}{*}{\# Params} & \multicolumn{4}{c}{Decoding method}    \\
 \cline{4-7}
                         &                          &                             & Greedy & Beam  & Attention   & Attention rescoring \\
                         \midrule
Transformer              & CTC AED                  & 29M                         & 34.55  & 34.4  & 40.61 & 32.15        \\
Conformer                & CTC AED                  & 31M                         & 28.73  & 28.72 & 31.60  & 27.38        \\
Conformer                & RNN-T AED                & 45M                         & 37.11  & 37.14 & 33.84 & 37.14        \\
Paraformer               & Paraformer               & 30M                         & 31.86  & 28.94 & -     & -           \\
\bottomrule
\end{tabular}
\caption{Decoding performance (CER, \%) of Transformer, Conformer, and Paraformer models trained from scratch}
\label{scratch}
\end{table*}

\begin{table*}[t]
\centering
\small
\begin{tabular}{cccccc}
\toprule
Model          & Architecture & Input   & \# Params & Sup./Self-sup.  & Training Data (hours) \\
\midrule
Wav2vec 2.0 (B)  & Enc            & Waveform       & 97M       & Self-sup. & 10K                   \\
Wav2vec 2.0 (L) & Enc            & Waveform       & 318M      & Self-sup. & 10K                   \\
HuBERT (B)     & Enc            & Waveform       & 97M       & Self-sup. & 10K                   \\
HuBERT (L)   & Enc           & Waveform       & 319M      & Self-sup. & 10K                   \\
CW      & Enc-Dec    & Fbank          & 122M       & Sup.      & 10K                   \\
Whisper        & Enc-Dec    & Waveform       & 39M-1,550M  & Sup.      & 680K                  \\
\bottomrule
\end{tabular}
\caption{Details of pre-trained baseline models. Enc and Dec stand for encoder and decoder, while Sup. and Self-sup. represent supervised and self-supervised learning. (B) and (L) denote the base and large versions.} 
\label{pretrain_baseline}
\end{table*}

\section{Tasks and baselines}
In this section, we evaluate our dataset on both ASR and SV tasks. 

\subsection{Speech recognition}

For child speech recognition, we trained several baseline models from scratch and fine-tuned pre-trained models to assess performance on our dataset. We use the Character Error Rate (CER, \%) as the evaluation metric. Refer to the Appendix \ref{Appendix_hyperparem} for the complete hyperparameter configurations.

\subsubsection{Baselines trained from scratch}
We utilize the open-source Wenet toolkit \citep{wenet} to train ASR models from scratch. Three architectures are chosen: Transformer \citep{transformer}, Conformer \citep{conformer}, and Paraformer \citep{paraformer}. These models incorporate different approaches, including Connectionist Temporal Classification (CTC) \citep{ctc}, RNN-Transducer (RNN-T) \citep{graves2012sequence}, and Attention-based encoder-decoder (AED) \citep{chorowski2014end,chan2015listen}.

The following models are considered:
\begin{itemize}
\item {\bf Transformer}: We trained the widely-used Transformer model with joint CTC/AED training. 
The training process follows the recipe and configuration provided by Wenet.
\item {\bf Conformer}: 
The Conformer \citep{conformer} integrates convolutions with self-attention for ASR.
We trained two models using both CTC and RNN-T loss functions respectively, following the Wenet recipe. 

\item {\bf Paraformer}: Proposed by Gao et al. \citep{paraformer}, Paraformer is a fast and accurate parallel transformer model.

\end{itemize}

\subsubsection{Results of training models from scratch}

Table \ref{scratch} presents the results of models trained from scratch on our dataset, evaluated using various decoding methods provided by Wenet \citep{wenet}. For Transformer and Conformer models with joint CTC and AED training \citep{kim2017joint}, we report CTC greedy and beam search decoding results. For Conformer models with RNN-T and attention loss, we include RNN-T greedy and beam search decoding results. All beam searches use a beam size of 10. 
Attention decoding and attention rescoring decoding results are also reported for Transformer and Conformer.

Conformer with CTC-AED performs best overall, achieving the lowest CER of 27.38\% with attention rescoring. Its CTC greedy and beam search methods yield nearly identical results (28.73\% and 28.72\%). In contrast, the Transformer model performs worse, with its best result being 32.15\% CER from attention rescoring, while Paraformer achieves competitive results, particularly with beam search (28.94\%).
RNN-T for Conformer performs less effectively, with no significant improvement from attention rescoring. Overall, Conformer with CTC-AED provides the most reliable performance, especially with attention rescoring.

\subsubsection{Pre-trained baselines}
We evaluate our dataset using a range of pre-trained baselines, including both supervised and self-supervised models. The details of these baselines are summarized in Table \ref{pretrain_baseline}.
For the supervised baselines, we include Conformer pre-trained on WenetSpeech \citep{wenetspeech} and Whisper \citep{whisper}. For the self-supervised models, we utilize Wav2vec 2.0 \citep{wav2vec2} and HuBERT \citep{hubert}, integrating a CTC decoder with the encoder to perform the ASR task.

\begin{itemize}

\item {\bf Wav2vec 2.0}: 
Wav2vec 2.0 \citep{wav2vec2} is a self-supervised model which jointly captures discrete speech units and contextualized features. 
We select two versions of Wav2vec 2.0 pre-trained using WenetSpeech.\footnote{\url{https://huggingface.co/TencentGameMate/chinese-wav2vec2-base} and \url{ https://huggingface.co/TencentGameMate/chinese-wav2vec2-large}}

\item {\bf HuBERT} :
HuBERT \citep{hubert} is a self-supervised model that uses k-means clustering to generate target labels and applies BERT-like prediction loss over masked audio regions to learn contextualized representations.
We select two versions of HuBERT pre-trained using WenetSpeech.\footnote{\url{https://huggingface.co/TencentGameMate/chinese-wav2vec2-large} and \url{https://huggingface.co/TencentGameMate/chinese-hubert-large}}

\item {\bf Conformer-WenetSpeech (CW)}: 
CW is a 122M-parameter Conformer CTC-AED model, trained with supervised learning on the WenetSpeech dataset. Checkpoints are available in Wenet’s open-source repository.\footnote{\url{https://github.com/wenet-e2e/wenet/blob/main/docs/pretrained_models.md}}

\item  {\bf Whisper}: 
Whisper \citep{whisper} is a Transformer-based multilingual ASR model trained on 680,000 hours of labeled speech data by OpenAI. We include various versions of Whisper, ranging from tiny to large, with model sizes from 39M to 1550M.\footnote{\url{https://github.com/openai/whisper}}
\end{itemize}

\begin{table}[] \small
\centering
\begin{tabular}{ccc}
\toprule
Model            & Greedy search & Beam search  \\
\midrule
Wav2vec 2.0 (B)  & 20.29         & 20.29       \\
Wav2vec 2.0 (L)  & 21.12         & 21.12            \\
HuBERT (B)    & 18.74         & 18.74       \\
HuBERT (L)    & 14.97         & 14.97       \\

\bottomrule
\end{tabular}
\caption{CER (\%) of self-supervised pre-trained baselines with greedy and beam search decoding}
\label{pretrain_res}
\end{table}

\begin{table}[] \small
\centering
\begin{tabular}{cccc}
\toprule
Model       & \# Params  & Zero-shot & Fine-tuning \\
\midrule
CW   & 122M  & 18.05     & 13.66   \\
\midrule
Whisper-tiny        & 39M   & 67.63     & 28.78      \\
Whisper-base        & 74M   & 51.49     & 23.33      \\
Whisper-small       & 244M  & 37.99     & 17.45      \\
Whisper-medium      & 769M  & 28.55     & 18.97      \\
Whisper-large-v2    & 1,550M & 29.43     & -          \\
\bottomrule
\end{tabular}
\caption{CER (\%) of supervised pre-trained baselines in zero-shot and fine-tuned settings}
\label{whisper_res}
\end{table}

\subsubsection{Results of fine-tuning pre-trained models}

Table \ref{pretrain_res} shows the CER for fine-tuning various self-supervised pre-trained models, including Wav2vec 2.0 and HuBERT, using both greedy and beam search decoding methods. HuBERT consistently outperforms Wav2vec 2.0, which is consistent with recent research \citep{superb}. 
Additionally, HuBERT (L) demonstrates better performance compared to its smaller counterpart, HuBERT (B). However, Wav2vec 2.0 (L) underperforms relative to Wav2vec 2.0 (B), likely due to overfitting, given the limited data size.

Table \ref{whisper_res} presents the CER results for Conformer-WenetSpeech (CW) and Whisper models in zero-shot and fine-tuning settings. Fine-tuning results in substantial CER improvements for all supervised models. Despite Whisper's large parameter size and extensive training data, the limited size of our dataset causes Whisper-medium to perform slightly worse than Whisper-Small after fine-tuning. Overall, CW achieves the best performance in both zero-shot and fine-tuned settings, highlighting its robust ASR capabilities learned from WenetSpeech.

\begin{figure}[!t]
  \centering
  \includegraphics[width=1.0\linewidth]
  {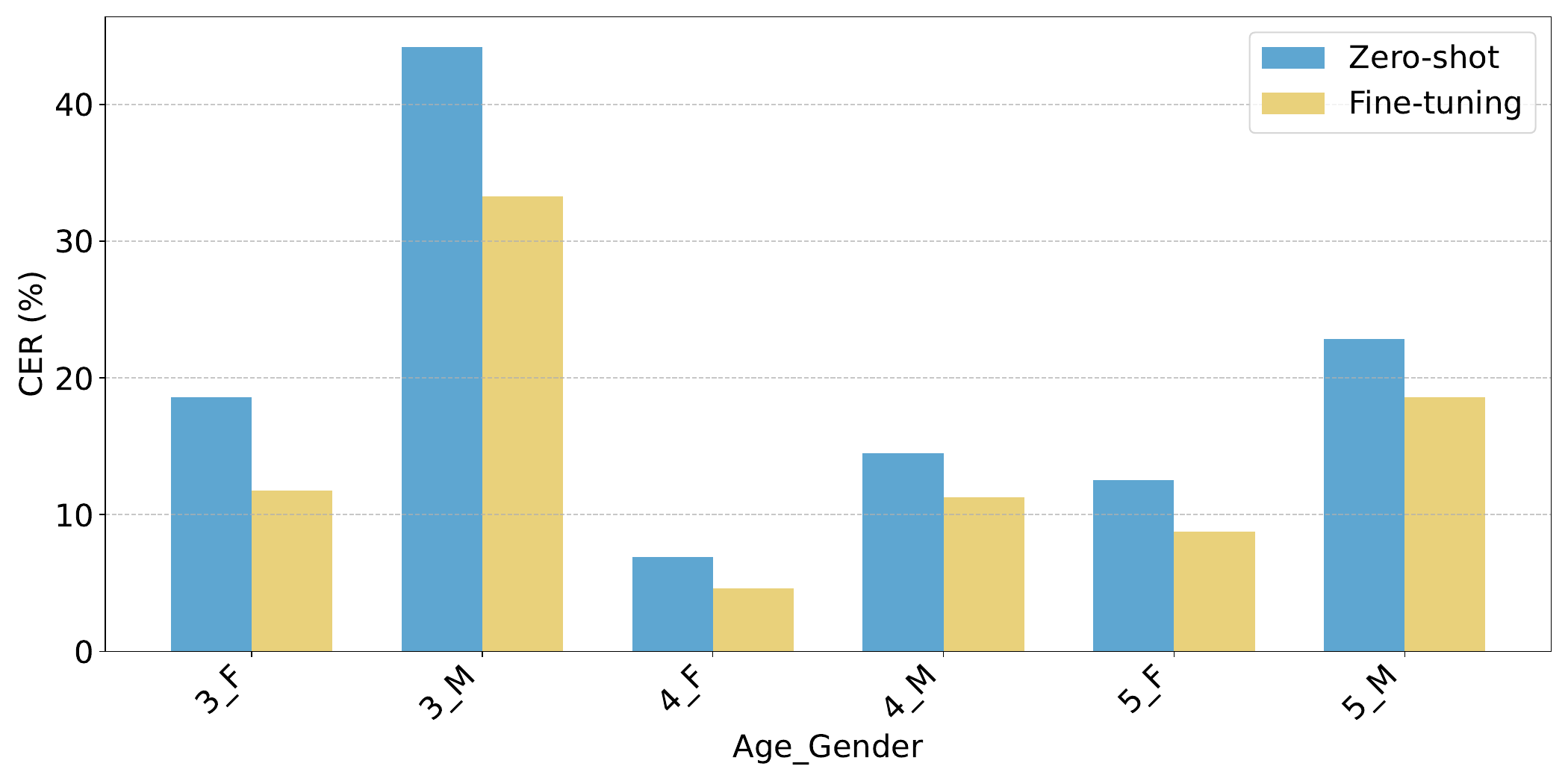}
  \caption{CER (\%) comparison of zero-shot and fine-tuning methods using CW model across different age-gender groups}
  \label{wer_comparison}
\end{figure}

\begin{table}[!t]
\tiny
\centering
\begin{tabular}{c c c c c c c}
\toprule
Method & \multicolumn{3}{c}{Zero-shot} &\multicolumn{3}{c}{Fine-tuning} \\ \midrule
\textbf{Age\_Gender} & S (\%) & D (\%) & I (\%) & S (\%) & D (\%) & I (\%) \\ \midrule
3\_F & 13.87 & 3.63 & 1.11 & 9.03  & 2.04 & 0.69 \\
3\_M & 34.78 & 5.45 & 3.97 & 26.80 & 4.35 & 2.11 \\
4\_F & 6.02  & 0.57 & 0.30 & 3.94  & 0.53 & 0.15 \\
4\_M & 11.55 & 1.80 & 1.14 & 8.92  & 1.46 & 0.91 \\
5\_F & 9.25  & 2.36 & 0.95 & 7.02  & 1.31 & 0.45 \\
5\_M & 16.70 & 4.36 & 1.81 & 14.32 & 3.04 & 1.23 \\
\bottomrule
\end{tabular}
\caption{In-depth comparison of different error types (S: Substitutions, D: Deletions, I: Insertions) between zero-shot and fine-tuning methods using CW model across different age-gender groups}
\label{scratch}
\end{table}

\begin{table*}[!t]
\centering
\small
  \resizebox{0.9\textwidth}{!}{
    \begin{tabular}{ccccccccc}
    \toprule
     \multirow{2}{*}{Model} & \multirow{2}{*}{\# Params} & \multirow{2}{*}{Dim} &\multirow{2}{*}{Dev (\%)} & \multicolumn{2}{c}{PLDA} & \multicolumn{2}{c}{Cosine similarity} \\ 
     \cmidrule(lr){5-6} \cmidrule(lr){7-8}
                    &       &                                &                               & EER (\%)        & minDCF       & EER (\%)              & minDCF              \\ 
                           \midrule
                           
    x-vector               &  4.2M & 512                            & 75.4                          & 8.91            & 0.7198         & 25.92                 & 0.9780                 \\
    ECAPA-TDNN         &   20.8M  & 192                            & 84.6                          & 13.72           & 0.8697         & 27.77                 & 0.9490                 \\
    ResNet-TDNN           &    15.5M  & 256                            & 91.9                          & 9.57            & 0.6597         & 22.11                 & 0.9044 \\
    \bottomrule
    \end{tabular}
  }
    \caption{Results of fine-tuning baselines on the speaker verification task, where Dim indicates the dimension of the extracted embeddings and Dev represents the accuracy on the validation set.}
  \label{sv}

\end{table*}

\subsubsection{Performance Analysis}
Figure \ref{wer_comparison} shows ASR performance across age and gender groups on the CW model. CER decreases with age, with 3-year-olds exhibiting higher error rates than 5-year-olds, reflecting greater variability in younger children's speech. Fine-tuning significantly reduces CER across all age groups, demonstrating its effectiveness in adapting models to children's speech.

\textbf{Gender Trends:}
Male speakers consistently exhibit higher CER than female speakers of the same age. This disparity may arise from greater pitch and articulation variability in young male children.

\textbf{Error Types:}
We further investigate error types in Tabel \ref{scratch}. Substitutions dominate error types, followed by deletions and insertions. Younger children, particularly 3-year-olds, exhibit higher substitution and deletion rates, reflecting challenges in speech recognition.

In summary, age and gender notably influence ASR performance, with younger and male speakers posing greater challenges. Fine-tuning mitigates these issues, highlighting the importance of targeted adaptation strategies. Detailed utterance analysis can be found in Appendix \ref{case_analysis}.

\subsection{Speaker verification}
In this section, we evaluate our dataset on the SV task. The evaluation is organized into three parts: dataset repartition, baselines, and results.
\subsubsection{Dataset repartition}
For the speaker verification task, the training and validation sets were merged, resulting in a total of 356 speakers. This combined data was then split into new training and validation sets with a 9:1 ratio for each speaker, while the test set remained unchanged. 
Although the training and validation sets share speakers, their speech samples are distinct. Verification trials were generated entirely from the test set, consisting of 20,000 trials and 41 speakers, with positive and negative trials evenly distributed (50\% each). The trials uniformly covered same-speaker pairs ${(spk_a, spk_a)}$ and different-speaker pairs ${(spk_a, spk_b)}$.

\subsubsection{Speaker verification baselines}
In this study, three popular speaker embedding extractors, pre-trained on VoxCeleb \citep{nagrani17_interspeech}, were fine-tuned on our dataset: x-vector\footnote{\url{https://huggingface.co/speechbrain/spkrec-xvect-voxceleb}} \citep{xvector2018}, ECAPA-TDNN\footnote{\url{https://huggingface.co/speechbrain/spkrec-ecapa-voxceleb}} \citep{ecapa2020}, and ResNet-TDNN\footnote{\url{https://huggingface.co/speechbrain/spkrec-resnet-voxceleb}} \citep{resnet2020}. These models were implemented using the SpeechBrain \citep{speechbrain} toolkit and fine-tuned for 40 epochs. The embeddings extracted from the verification trials were then used to evaluate the models' performance on the SV task. Refer to the Appendix \ref{Appendix_hyperparem} for the complete hyperparameter configurations.

\subsubsection{Results of speaker verification}
For evaluation, two scoring methods were applied: Probabilistic Linear Discriminant Analysis (PLDA) \citep{prince2007probabilistic} and Cosine Similarity. Performance was measured using two metrics: Equal Error Rate (EER) and Minimum Detection Cost Function (minDCF). EER is computed by finding the verification threshold where the false rejection and false acceptance rates ($p_{miss}$ and $p_{fa}$) are equal, such that EER $= p_{fa} = p_{miss}$. The DCF is calculated using:
$$
C_{\delta} = c_{miss} \cdot p_{miss} \cdot p_{target} + c_{fa} \cdot p_{fa} \cdot (1 - p_{target})
$$
where $c_{miss}$ is the cost of false rejection, $c_{fa}$ is the cost of false acceptance, and $p_{target}$ represents the probability that the target speaker appears in the verification set. In this case, $c_{miss} = c_{fa} = 1 $ and $p_{target} = 10^{-2}$.

Table \ref{sv} summarizes the performance of the models on the dataset, with both PLDA and Cosine Similarity evaluated using EER and minDCF metrics. Two key insights emerge from the results:
First, the dataset proves to be well-suited for speaker-related tasks, as indicated by the strong performance of the three fine-tuned baseline models. However, the underdeveloped vocal characteristics of young children present challenges, potentially masking gender-related features and other distinguishing attributes.
Second, due to the relatively small size of the dataset, the larger ECAPA-TDNN model underperformed compared to ResNet and x-vector, likely due to overfitting. 

\section{Conclusion}

In conclusion, this paper introduces a valuable Mandarin speech dataset specifically designed for young children aged 3 to 5, addressing a crucial gap in ASR resources for this age group. Comprising 41.25 hours of speech data from 397 speakers across diverse provinces in China, the dataset ensures balanced gender representation and board geographic coverage. Our evaluations of ASR models and speaker verification show significant improvements, highlighting the dataset's effectiveness in advancing children's speech technology. This work represents a significant contribution to Mandarin child speech research and holds great promise for applications in educational technology and child-computer interaction.

\section*{Limitations}
Despite the dataset comprising 41.25 hours of speech data, it remains relatively small compared to adult speech datasets, which typically encompass much larger volumes. Additionally, while the dataset covers 22 provinces across China, the geographic distribution is not fully balanced, and expanding representation from underrepresented regions could improve diversity.
Overfitting can occur when fine-tuning pre-trained models with a large number of parameters, particularly on smaller datasets. To address this, parameter-efficient fine-tuning methods like LoRA \citep{hu2022lora} could be explored to enhance model performance.

\section*{Ethics Statement}
This study adhered to strict ethical standards to safeguard the well-being and rights of participants. Recordings were conducted in a conversational context to encourage natural interaction, with parents present to provide emotional support. The content was unrestricted, focusing on age-appropriate, familiar communication to ensure a stress-free environment.

Informed consent was obtained from parents or legal guardians, who were fully briefed on the study's purpose, procedures, and data use for academic research. They were informed of their right to withdraw at any time. Each child received a fair compensation of 150 RMB (about \$20 USD), carefully calibrated to avoid undue influence.

To protect privacy, all data was anonymized, removing personal identifiers and replacing them with coded labels. The dataset is securely stored, with access restricted to authorized researchers for academic purposes. The publicly available dataset will be licensed to prohibit commercial use and ensure compliance with ethical research practices.

While the study posed minimal risks, measures such as parental presence and familiar settings were implemented to ensure children’s psychological comfort. This dataset aims to advance automatic speech recognition (ASR) technologies for underrepresented groups like young children. However, we recognize the potential misuse of ASR technologies and have taken steps to mitigate risks by restricting dataset use to academic research and promoting ethical applications.

In summary, this study emphasizes informed consent, privacy protection, fair compensation, and ethical use of the data, ensuring respect for participants' rights and well-being while contributing responsibly to the research community.

\bibliography{custom}

\appendix
\section{Experimental configurations} \label{Appendix_hyperparem}

\setcounter{figure}{0}  
\setcounter{table}{0}   
\renewcommand{\thefigure}{A.\arabic{figure}}
\renewcommand{\thetable}{A.\arabic{table}}

\begin{table*}[!t]
\centering
\begin{tabular}{cccccc}
\toprule
\textbf{Encoder}     & \textbf{Decoder}    & \textbf{Batch size} & \textbf{LR} & \textbf{Warmup} & \textbf{Epochs} \\ 
\midrule
Transformer & CTC+ATT    & 32         & 1.00E-03      & 2500         & 100              \\
Conformer   & CTC+ATT    & 32         & 1.00E-03      & 2500         & 100              \\
Conformer   & RNN-T+ATT  & 32         & 1.00E-03      & 2500         & 100              \\
Paraformer  & Paraformer & 16         & 2.00E-03      & 2500         & 100    \\
\bottomrule         
\end{tabular}
\caption{Hyperparameters for training ASR models from scratch.}
\label{table:asr_scratch}
\end{table*}

\begin{table*}[!t]
\centering
\begin{tabular}{cccc}
\toprule
\textbf{Model}           & \textbf{Batch size} & \textbf{Learning rate}         & \textbf{Epochs} \\
\midrule
CW              & 16         & 4.00E-04              & 100              \\
Wav2vec 2.0 (B) & 10         & 3.00E-05              & 70               \\
Wav2vec 2.0 (L) & 10         & 1.00E-04              & 70               \\
HuBERT (B)      & 10         & 5.00E-05              & 70               \\
HuBERT (L)      & 10         & 5.00E-05              & 70               \\
Whisper         & 16         & 1.00E-5 $\sim$ 1.00E-6 & 20    \\          
\bottomrule         
\end{tabular}
\caption{Hyperparameters for fine-tuning pre-trained ASR models.}
\label{table:asr_finetuning}
\end{table*}

\begin{table*}[!t]
\centering
\begin{tabular}{cccccc}
\toprule
\textbf{Model}       & \textbf{Batch size} & \textbf{LR Schedule} & \textbf{Init LR}  & \textbf{Base LR} & \textbf{Epochs} \\ 
\midrule
ECAPA-TDNN  & 128        & Cyclic      & 5.00E-03 & 1.00E-08 & 40               \\
ResNet-TDNN & 128        & Cyclic      & 5.00E-03 & 1.00E-08 & 40               \\
X-vector     & 128        & Linear      & 5.00E-03 & 1.00E-04 & 40              
\\          
\bottomrule         
\end{tabular}
\caption{Hyperparameters for training speaker verification models.}
\label{table:sv_training}
\end{table*}

This section provides detailed configurations and hyperparameters used for training and fine-tuning the ASR and speaker verification (SV) models discussed in the paper. All experiments were conducted using four GTX 3090 or GTX 4090 GPUs over several hours.

\subsection{ASR model training from scratch}
The hyperparameters for training ASR models from scratch are summarized in Table \ref{table:asr_scratch}. These models were trained using the Wenet toolkit with the configurations shown below.

\subsection{ASR model fine-tuning}
Table \ref{table:asr_finetuning} presents the fine-tuning hyperparameters for pre-trained ASR models, including Wav2vec 2.0, HuBERT, Whisper, and Conformer-WenetSpeech. Fine-tuning was performed using the training subset of our dataset.

\subsection{Speaker verification (SV) model training}
Table \ref{table:sv_training} provides the training configurations for speaker verification models, including ECAPA-TDNN, ResNet-TDNN, and X-vector. These models were trained and evaluated on our dataset for speaker verification tasks.

\section{Analysis of fine-Tuning performance on specific utterances} \label{case_analysis}

\setcounter{figure}{0}  
\setcounter{table}{0}   
\renewcommand{\thefigure}{B.\arabic{figure}}
\renewcommand{\thetable}{B.\arabic{table}}

\begin{figure*}[!t]
  \centering
  \includegraphics[width=1.0\linewidth]
  {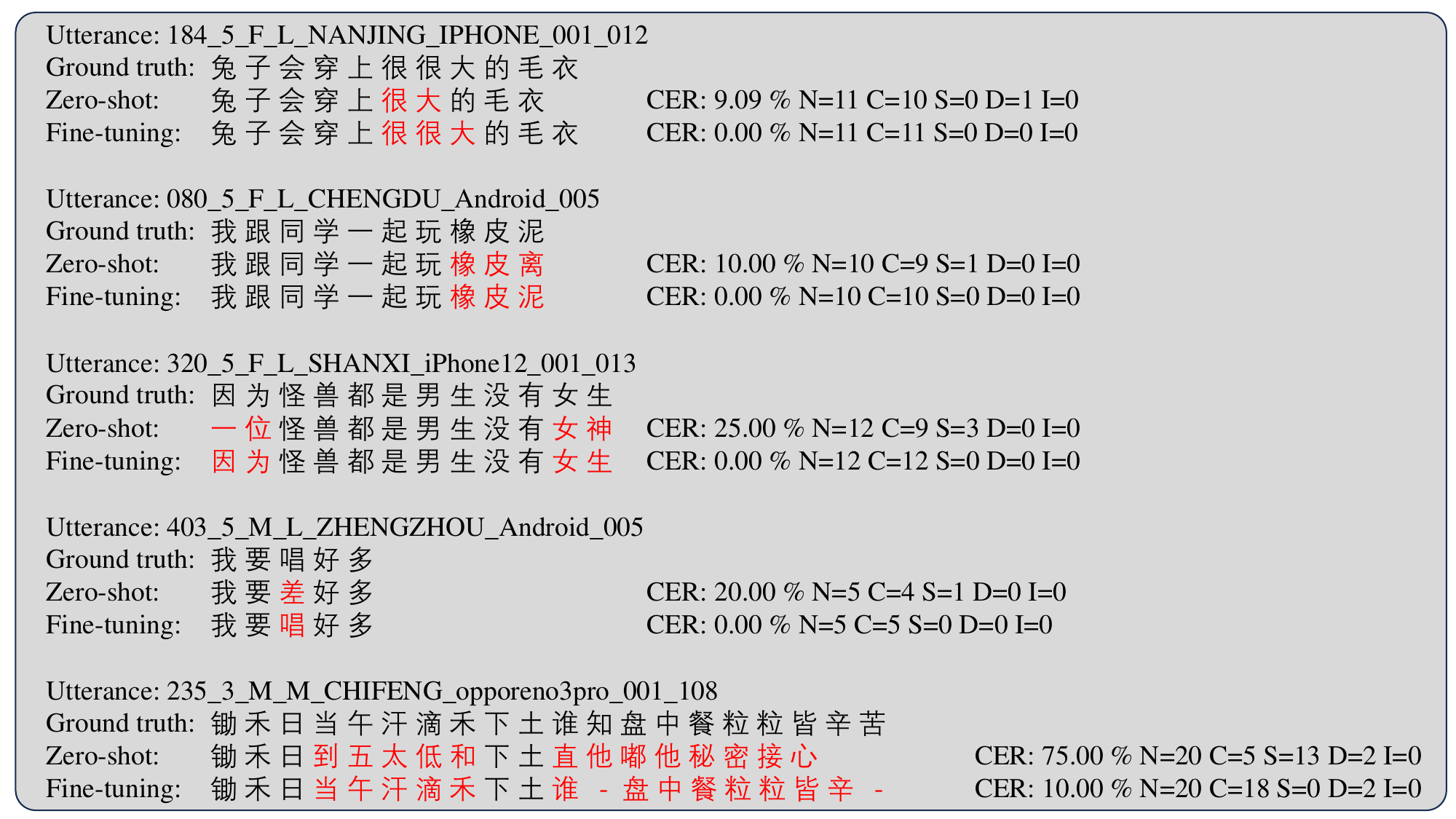}
  \caption{Performance comparison of zero-shot and fine-tuned models on specific utterances}
  \label{some_case}
\end{figure*}

As presented in Figure \ref{some_case}, the fine-tuning process significantly improved the ASR model's performance across various utterances, with a clear reduction in character error rate (CER).
In general, fine-tuning allowed the model to adapt to specific child speech variations, addressing common issues such as phoneme substitutions and mispronunciations. Despite these improvements, some residual errors were still observed, particularly for more complex or longer utterances. Overall, the results demonstrate the effectiveness of fine-tuning for enhancing ASR performance on child speech, though further optimization is necessary to fully address all challenges.


\end{document}